\documentclass[pre,showpacs,preprint]{revtex4}

\usepackage{amsmath}

\usepackage{graphicx}
\usepackage{dcolumn}
\usepackage{bm}

\begin{document}

\title{Volume fluctuations and compressibility of a vibrated granular gas}
\author{J. Javier Brey and M.J. Ruiz-Montero}
\affiliation{F\'{\i}sica Te\'{o}rica, Universidad de Sevilla,
Apartado de Correos 1065, E-41080, Sevilla, Spain}
\date{\today }

\begin{abstract}
The volume fluctuations in the steady state reached by a vibrated granular gas of hard particles 
confined by a movable piston on the top
are investigated by means of event driven simulations. Also, a compressibility factor, measuring
the response in volume of the system to a change in the mass of the piston, is introduced and measured.
From the second moment of the volume fluctuations and the compressibility factor, an effective temperature
is defined, by using the same relation as obeyed by equilibrium molecular systems. The interpretation of
this effective temperature and its relationship with the granular temperature of the gas, and also with the velocity
fluctuations of the movable piston, is discussed. It is found that the ratio of the temperature based on the volume fluctuations
to the temperature based on the piston kinetic energy, obeys simple dependencies on the inelasticity and on
the piston-particle mass ratio.

\end{abstract}


\maketitle

\section{Introduction}
\label{s1}
Granular matter in general \cite{JNyB96,Du00}, and granular gases in particular \cite{Go03,ByP04,AyT06},
have been recently the object of intensive theoretical and experimental research. Not in the least this
is because of the opportunity they offer to investigate fundamental questions of non-equilibrium physics.
One of these issues is the nature and properties of non-equilibrium fluctuations and their relevance for
the description of the macroscopic behavior of the system. The simplest fluctuations that can be considered
are those of global properties of a system. The total energy fluctuations of an isolated
granular gas, modeled as an ensemble of smooth inelastic hard spheres, have already been  investigated
elsewhere \cite{BGMyR04,VPBWyT06}. They are specific of granular systems,  and  a consequence of the localized
character of the energy dissipation in collisions.

In this paper, the volume fluctuations of a vibrated granular gas are
studied by means of event driven simulations \cite{AyT87,PyS05}. The volume of the system changes because the wall on
top of the grains is a piston that can move in the vertical direction. The equilibrium position of the piston
is determined by equating its weight per unit of area with the pressure of the granular gas just below it.
On the other hand, the properties of the fluctuations of the piston around that position, and therefore
the volume fluctuations of the system, are not known. If the system were a molecular gas at equilibrium,
therefore without being energized by vibration, the total volume fluctuations are known to be Gaussian, and with
a second moment that is proportional to the temperature and the isothermal compressibility of the 
gas \cite{LyL79}. There is no
reason to expect the above properties to hold in granular gases, that are inherently in non-equilibrium states. Even
more, to give  a meaning to them, a first point to be addressed is to specify the notion of temperature to be used.
Actually, the equilibrium
expression for the volume fluctuations can be used to define a temperature-like quantity, that can be experimentally
measured. This is somehow an extension of the usual way of defining ``effective temperatures'' from the extension to
non-equilibrium states of the Fluctuation-Dissipation Theorem. A revision of these ideas in the context of
granular media is given in \cite{BBALMyP05}. See also \cite{PByV07} for a discussion of the validity of the Einstein
relation in externally driven granular gases.

The emerging natural question is whether the effective temperature defined
from the relationship between volume fluctuations and compressibility has some other conceptual interpretation,
and if it is related to other possible sensible definitions of temperature in the system. The first obvious
candidate to be considered is the granular temperature of the gas, defined from the second moment of the
velocity distribution of the grains. Actually, this parameter is known to play in the hydrodynamic description of
granular gases a role similar to the usual temperature in molecular hydrodynamics \cite{Ca90}. But there is another
temperature parameter that is relevant for the description of the system, the one defined from the second moment of
the velocity distribution of the movable piston. Both temperatures, defined from the velocity of the gas and of the piston, respectively,
are the same in equilibrium systems, but they can differ strongly in granular systems, as a manifestation of
the violation of energy equipartition \cite{ByR08a}. Clarifying the relationship between the effective temperature
defined from the volume fluctuations, the granular temperature, and the temperature parameter of the piston, is
one of the aims of this work.

The structure  of the remaining of this paper is as follows. In Sec. \ref{s2}, the system is described
and the macroscopic steady state to be considered is characterized. Also, the parameter region to be
investigated is specified. It is shown than when the mass of the piston is much larger than the mass of the
grains, the volume fluctuations of the system exhibit a Gaussian distribution. The second moment of this distribution
depends much stronger on the inelasticity of collisions between particles than on the elastic or inelastic
character of the collisions of the particles with the piston. In Sec. \ref{s3}, a compressibility factor is defined for the
granular gas. As usual, it is a measurement of the change in the volume of the system as a consequence of a change in
the external pressure under well defined constrains. From the values of  the second moment of the volume fluctuations and of the compressibility factor,
an effective temperature is defined, as indicated above. This temperature turns out to be quite simply related with
the temperature parameter of the piston, but its relationship with the granular temperature of the gas
appears to be very intricate. Finally, Sec. \ref{s4} contains a short summary of the main results in the paper and some
general comments.

\section{Steady fluctuations of the position of the piston}
\label{s2}
Consider a system composed by $N$ smooth inelastic hard disks of mass $m$ and
diameter $d$, in presence of gravity, and confined by a movable piston of mass $M$
located on the top. By definition, the piston can only move in the direction of the
gravity field. Moreover, there is no friction between the piston and the lateral
walls of the vessel containing the gas. The system of particles is kept fluidized and at low density
by injecting energy through the
bottom wall, which is vibrating. Inelasticity in collisions between particles is modeled by means of a
constant coefficient of normal restitution,
$\alpha$, defined in the interval $0 < \alpha < 1$. Thus when two particles $i$ and $j$ collide
their velocities change instantaneously from the initial values ${\bm v}_{i}$, ${\bm v}_{j}$ to the
post-collisional ones given by
\begin{equation}
\label{2.1}
{\bm v}_{i}^{\prime} = {\bm v}_{i}- \frac{1+ \alpha}{2} \left( \widehat{\bm \sigma} \cdot {\bm v}_{ij}
\right) \widehat{\bm \sigma},
\end{equation}
\begin{equation}
\label{2.2}
{\bm v}_{j}^{\prime} = {\bm v}_{j} + \frac{1+ \alpha}{2} \left( \widehat{\bm \sigma} \cdot {\bm v}_{ij}
\right) \widehat{\bm \sigma},
\end{equation}
where ${\bm v}_{ij} \equiv {\bm v}_{i}-{\bm v}_{j}$ is the relative velocity and $\widehat{\bm \sigma}$ is
the unit vector joining the center of the two particles at contact. The $z$ axis will be taken in the direction
of the gravitational field, so that the particles are submitted to an external force of the form ${\bm f}=- m g_{0}
\widehat{\bm e}_{z}$, $g_{0}$ being a positive constant and $\widehat{\bm e}_{z}$ the positive unit vector
along the $z$ axis. Collisions of particles with the movable piston on the top are also considered as smooth and 
inelastic, $\alpha_{P}$ being the coefficient of normal restitution for them. 
Therefore, in a collision between particle $i$ and the piston,
the component $v_{i,x}$ of the velocity of the particle perpendicular to the $z$ axis remains unchanged,
\begin{equation}
\label{2.3} v_{i,x}^{\prime}= v_{i,x},
\end{equation}
while the component $v_{i,z}$ of the velocity of the particle and the velocity
$V_{z}$ of the piston are instantaneously modified accordingly with
\begin{equation}
\label{2.4} v^{\prime}_{i,z} =v_{i,z} -\frac{M}{m+M}\, (1+\alpha_{P})
(v_{i,z}-V_{z})
\end{equation}
and
\begin{equation}
\label{2.5} V^{\prime}_{z} =V_{z} +\frac{m}{m+M}\, (1+\alpha_{P}) (v_{i,z}-V_{z}),
\end{equation}
respectively. When the transversal section of the system, i.e. the size of the piston $W$, is smaller than
a critical value \cite{BRMyG02,LMyS02}, the gas reaches, after a transient time interval, a stationary state with gradients only in the
direction of the gravitational field. If, in addition, the inelasticity of the system is small, the inelastic hydrodynamic
Navier-Stokes equations with the appropriate boundary conditions \cite{BDKyS98,ByC01} provide an accurate description of
the stationary state \cite{ByR09a}. As the coefficient of normal
restitution of the gas $\alpha$ decreases, significant deviations from the predictions following from
the Navier-Stokes equations show up. They are due to the coupling between inelasticity and gradients that
exists in the stationary state, in such a way that strong inelasticity implies large gradients of the hydrodynamic fields.
This coupling is a peculiarity of the steady states of inelastic fluids, following from the balance between the energy
dissipated because of inelastic cooling and the hydrodynamic energy fluxes.

A previous analysis, carried out in \cite{ByR09a}, focussed on the macroscopic description of the granular gas
in terms of the density and granular
temperature fields, the velocity field being zero. At this level of description, the role of the movable piston
on top of the gas is to partially determine  the boundary conditions needed to solve the hydrodynamic equations for
the steady state under consideration. Here, the interest will be on the fluctuations of the movable piston, namely on its
position fluctuations. Some results for the velocity fluctuations have been reported elsewhere \cite{ByR08a}.
There, it was shown that the steady state velocity fluctuations of the piston are gaussian with zero mean for 
$\alpha_{P} \geq 0.6$ and $\alpha \geq 0.8$. Nevertheless, no simple relationship between the second moments
of the velocity distributions of the piston and the gas next to it was found. It is worth to remark that
there is no reason to expect such a relation to exist at a macroscopic level of description,
i.e. involving only the hydrodynamic fields and the parameters of the system. Actually, the simulation results
reported in \cite{ByR08a} indicate that the details of the velocity distribution of the gas, beyond its first 
few moments, are relevant to determine the second moment of the velocity distribution of the piston.

It is clear that the position fluctuations of the movable piston are related with the volume
fluctuations of the inelastic gas. Actually, this relationship can be made direct and exact by properly choosing
the nature of the vibrating wall located at the bottom of the system. The mission of the latter is to energize the system,
keeping the particles fluidized. The expectation is that the behavior in the bulk of the system is independent of the details of the
way in which this wall is being vibrated. Consequently, the
simplest possible choice has been used in all the results to be reported in the following.
The bottom wall is vibrated with a sawtooth velocity profile, having a velocity $v_{W}$. This means that all the
particles colliding with the wall find it moving upwards with that velocity \cite{McyB97,McyL98}. Besides, the amplitude of
the wall motion is considered much smaller than the mean free path of the particles in its vicinity, so that
the position of the wall can be taken as fixed at $z=0$. Therefore, the dynamics of the
vibrating wall at the bottom does not induce directly any change in the volume (or area) occupied by the granular gas.
Also, and again for the sake of simplicity, collisions of the particles with this wall will be considered as elastic.

In the event-driven simulations carried out, periodic boundary conditions were
used in the $x$ direction. The width of the system and the number of particles in it were fixed to $W= 70 d$
and $N_{z} \equiv N / W =6 d^{-1}$, respectively. Moreover, the velocity of the
vibrating wall $v_{W}$ was chosen in each case large enough, not only to fluidize the system, but also
to guarantee that the density remains small throughout the granular gas and, consequently, the dilute limit can be expected to
be accurate. In this case, the dependence of the hydrodynamic profiles on $v_{W}$ is very simple
and follows by dimensional analysis \cite{ByR09a}. The dependence of the position fluctuations of the piston on this velocity
will be discussed later on. The value of the coefficient of normal restitution for the collisions
between particles has been varied within the interval $0.85 \leq \alpha <1$. This includes a range of values for
which the Navier-Stokes hydrodynamic description is not accurate, due to the coupling between inelasticity and
gradients already pointed out \cite{ByR09a}. The coefficient of restitution for the collisions of the particles with the
movable piston has been set to $\alpha_{P}=0.99$ in most of the simulations being reported, but it has been verified that
the results depend very
weakly on the value of this coefficient, remaining practically the same when $\alpha_{P}$ is decreased, at least
up to $\alpha_{P}=0.8$. Some examples of this behavior will be given below.

\begin{figure}
\includegraphics[scale=0.7,angle=0]{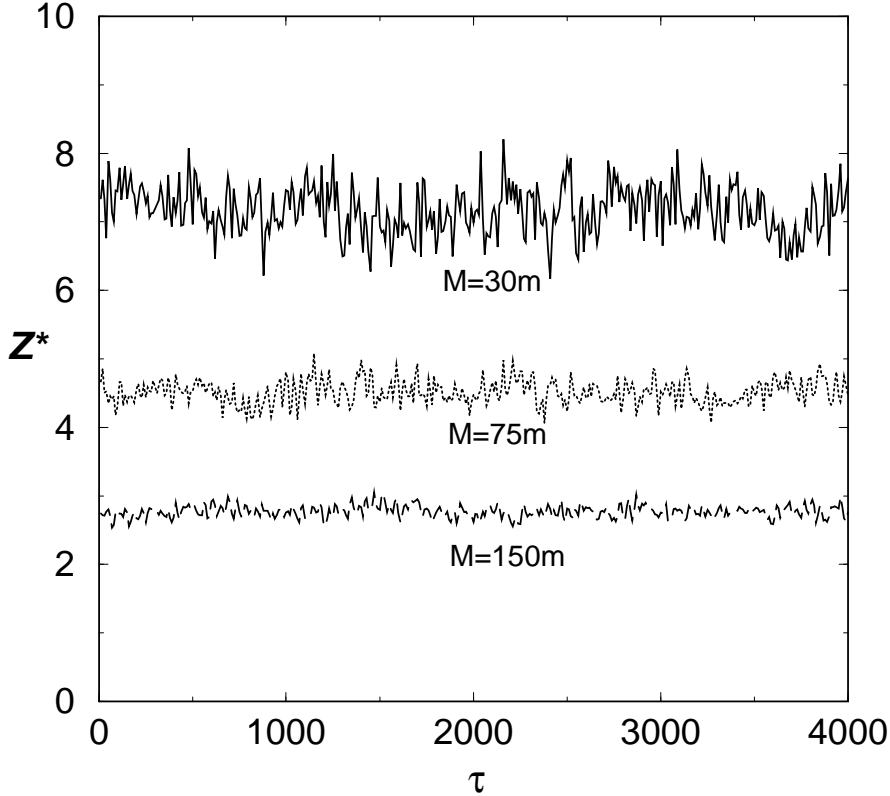}
\caption{Time evolution of the dimensionless scaled position $Z^{*} \equiv Z g_{0}/ v_{W}^{2}$ of the movable piston
located on top of the system, once the steady state has been reached. There are $420$ disks in a box of width
$W=70 d$. In all cases the coefficient of normal restitution of the particle collisions is $\alpha= 0.95$ and that
for the particle-piston collisions is $\alpha_{P}=0.99$. Results for three
values of the mass of the piston M are shown, as indicated. Time $\tau$ is measured in accumulated number of collisions
per particle. } \label{fig1}
\end{figure}

In all the simulations being presented, it was observed that the height $Z$ of the movable piston oscillates about
an average value $<Z> =L$, once the steady state is reached.  As an example, in Fig.\  \ref{fig1} the time evolution
of the scaled position of the piston, $Z^{*} \equiv Z g_{0} /v_{W}^{2}$, is shown for $\alpha = 0.95$, $\alpha_{P}=0.99$,
and three
choices of the mass of the piston: $M=30 m$, $M=75 m$, and $M=150m$. Time $\tau$ is measured in accumulated
number of collisions per particle. It is observed that both, the value of $L$ and the amplitude of the fluctuations,
decrease as $M$ increases. Of course, this is the expected behavior. Also notice that the trajectory
of the piston does not exhibit systematic oscillations, but it is apparently random. This indicates that the motion
of the piston does not have any hydrodynamic component induced, for instance, by the vibrating wall at the bottom.

\begin{figure}
\includegraphics[scale=0.7,angle=0]{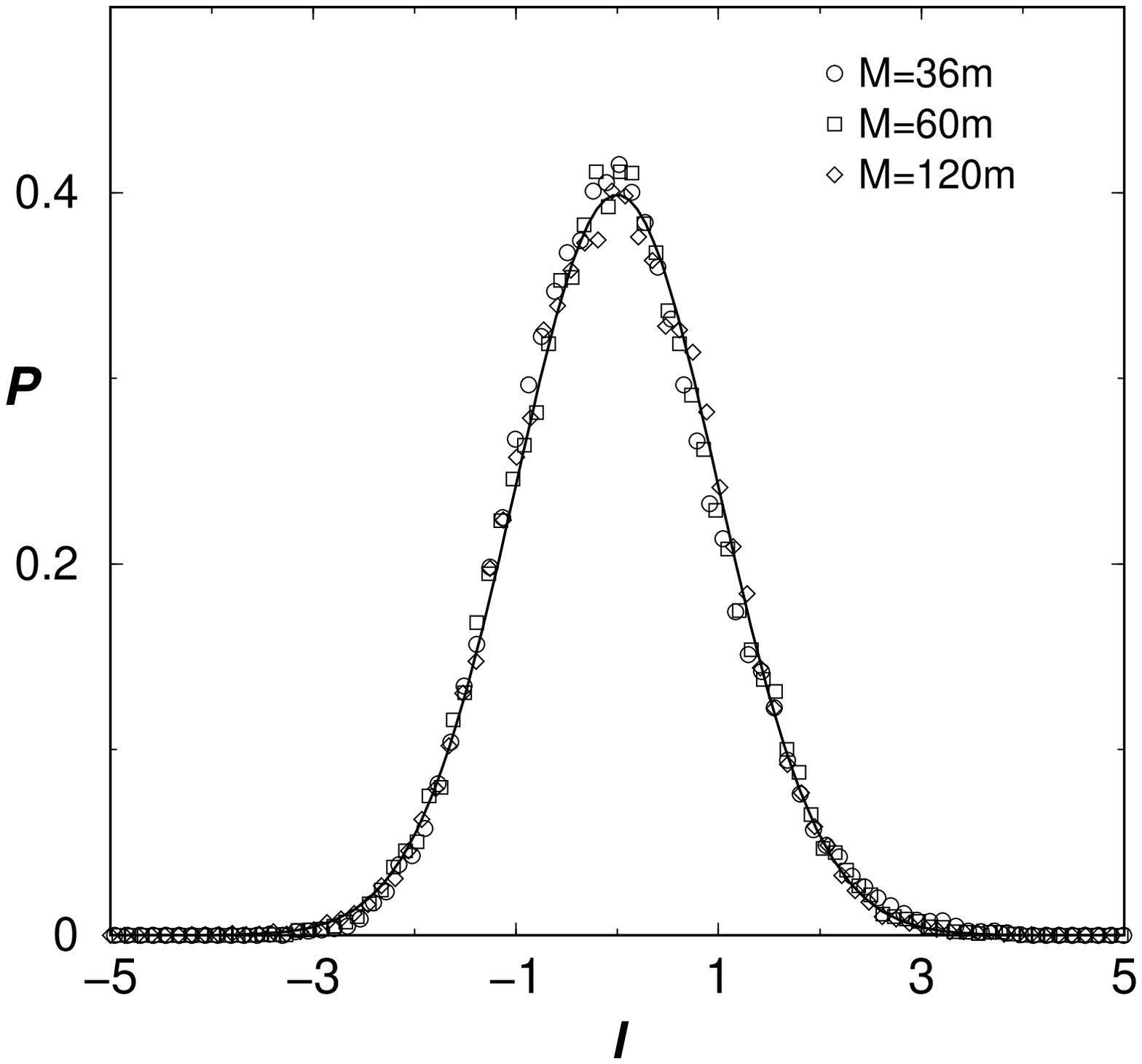}
\includegraphics[scale=0.7,angle=0]{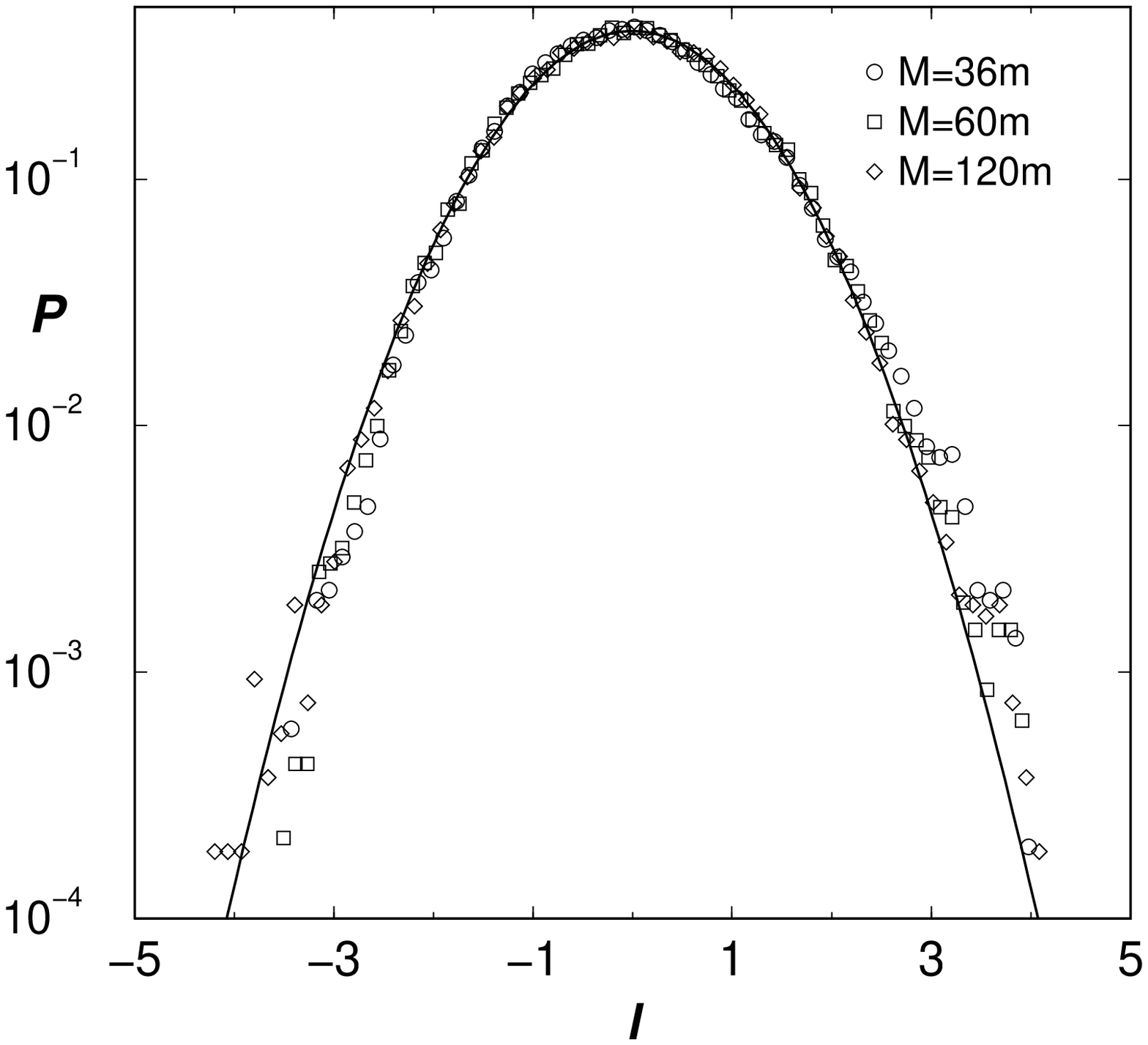}
\caption{Steady position distribution of the piston $P(l)$ in both normal and logarithmic scales. The symbols are
from the simulations while the solid lines are Gaussian with unity dispersion. The data correspond to three systems differing
in the mass of the piston, as indicated. The coefficient of normal restitution for the particle collisions is $\alpha=0.94$.
The dimensionless length $\ell$ is defined in Eq.\ (\protect{\ref{2.6}}).} \label{fig2}
\end{figure}

From the steady trajectory of the piston, the probability distribution for its position can be built. To increase
the statistics, several trajectories have been generated for each set of values of the parameters.  As already mentioned,
the fluctuations of the piston depend on the parameters defining the system. To see whether this dependence occurs
only through the first two moments of the probability distribution, a normalized length $\ell$ has been defined as
\begin{equation}
\label{2.6}
\ell  \equiv \frac{Z-L}{\sigma_{Z}}=\frac{Z^{*}-L^{*}}{\sigma_{Z}^{*}},
\end{equation}
where $\sigma_{Z}$ is the square root of the second central moment or standard deviation of $Z$, $\sigma_{Z}^{2} \equiv
<Z^{2}>-L^{2}$, and the star indicates that lengths are being measured in
the dimensionless scale defined above. In Fig. \ref{fig2}, the obtained probability distribution of $\ell$,
$P(\ell)$, is plotted for a system with $\alpha=0.94$. Again, three values of the mass ratio have been considered, namely
$M=36 m$, $M = 60 m$, and $M= 120 m$. It is seen that the probability distributions are accurately fitted by a Gaussian
(solid lines), at least up to values of the probability density of the order of $10^{-4}$. A similar behavior has been
found in all the simulated systems with parameters within the ranges mentioned above, although it seems that a
small but systematic deviation shows up as the mass of the piston $M$ becomes smaller, approaching the mass of the
particles. A possible explanation for this behavior is that, as the mass of the piston decreases, the amplitude of its position
fluctuations increases, and the effect of the external gravitational field breaks the symmetry of the fluctuations around the
average position. To check this idea and to quantify the deviations from the Gaussian of
the position fluctuations, the third and fourth moments
of $\ell$ have been computed from the simulation data. The results for some of the simulations are given in Table
\ref{table1}. For a Gaussian
distribution it is $<\ell^{3}> = 0$ and $<\ell^{4}> =3$. The deviations of the third moment from the Gaussian value are much stronger than those
of the fourth one, supporting the idea that the main cause of the deviation from the Gaussian is due to the symmetry breaking
produced by the external field, when the mass of the piston is not much larger than the mass of the particles. In any case,
the deviations are rather weak and it can be concluded that, in the explored parameter region, the position fluctuations
of the piston can be considered as Gaussian with a very good accuracy.

\begin{table}
\caption{Third and fourth moments of the position distribution of the piston, obtained from the  simulations.
The position $\ell$ is given in the dimensionless scale defined in Eq.\ (\protect{\ref{2.6}}).} \label{table1}
\begin{ruledtabular}
\begin{tabular}{lllll}
$ \alpha $ & $ \alpha_{P} $ & $ M/m $ & $ <\ell^{3}> $  & $ <\ell^{4}> $ \\
0.98 & 0.99 & 24 & 0.250 & 3.083 \\
     &      & 60 & 0.172 & 3.190 \\
     &      & 120 & 0.0385 & 2.898 \\
     & 0.8  & 24 & 0.254 & 3.230 \\
     &      & 60 & 0.136 & 3.047 \\
     &      & 120 & 0.057 & 2.922 \\
0.94 & 0.99 & 24 & 0.242& 3.138 \\
     &      & 60 & 0.125 & 2.996 \\
     &      & 120 & 0.047 & 3.006 \\
     & 0.8  & 24 & 0.145 & 2.964 \\
     &      & 60 & 0.213 & 3.088 \\
     &      & 120 & 0.101 & 2.985 \\
\end{tabular}
\end{ruledtabular}
\end{table}

In ref. \cite{ByR09a} it was shown that the values of the average position of the piston $L$ scale with $v_{W}^{2}$ in the
low density limit and, consequently, $L^{*}$ does not depend on the velocity of the vibrating wall in this limit.
The extension of this result to volume fluctuations requires to go beyond hydrodynamics.
To investigate whether $\sigma^{*}_{Z}$ also has the above scaling property, several series of simulations have been performed varying
the value of $v_{W}$, while keeping constant all the other parameters. It is important to stress that
$v_{W}$ was always chosen large enough as to fluidize the system and to avoid the presence of regions with density above
what is considered the low density range. In all the cases investigated, there was no dependence of $\sigma^{*}_{Z}$ on
$v_{W}$,  within the
statistical uncertainties. Therefore, in the low density limit, the standard deviation of the piston position seems to
scale with the  square of the velocity of the vibrating wall, i.e. in the same  way as the average position $L$.

\begin{figure}
\includegraphics[scale=0.7,angle=0]{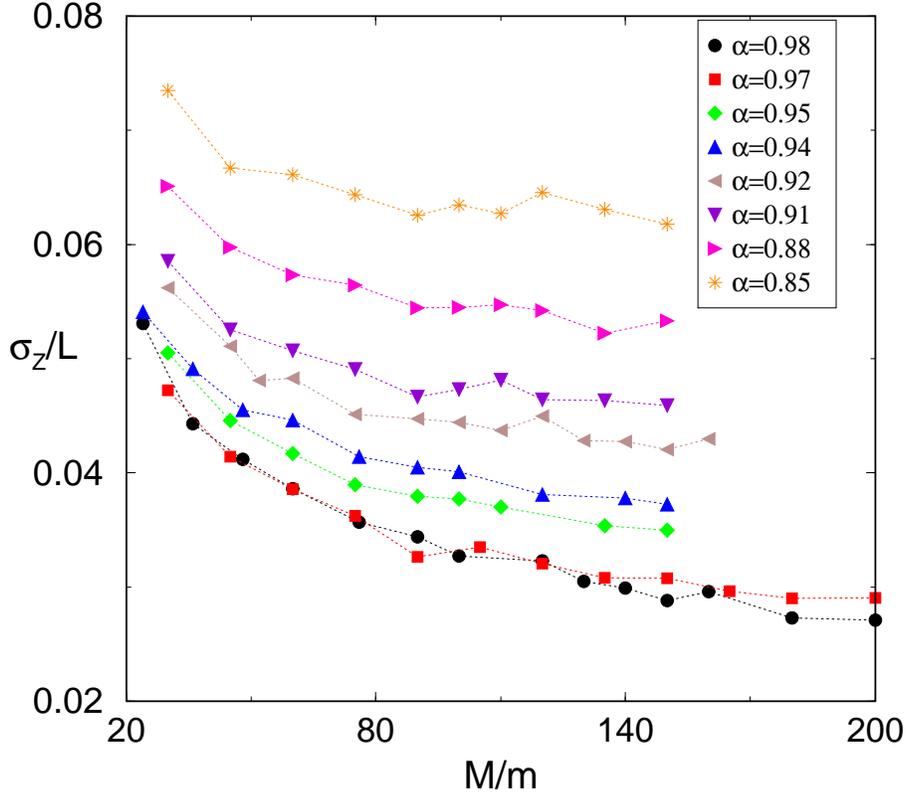}
\caption{(Color online) Dimensionless relative standard deviation $\sigma_{Z}/L$ of the position of the
movable piston as a function of the mass ratio $M/m$, for several values of the coefficient of normal restitution of
the gas $\alpha$, as indicated in the insert. The curves are guides for the eye. In all the cases, the coefficient of restitution
for the gas-movable piston collisions is $\alpha_{P}=0.99$.}
\label{fig3}
\end{figure}

In Fig.\ \ref{fig3}, the relative standard deviation $\sigma_{Z} / L = \sigma_{Z}^{*} / L^{*}$ is plotted as a function of
the mass ratio $M/m$, for several values of the coefficient of normal restitution of the gas $\alpha$ in the interval
$0.85 \leq \alpha \leq 0.98$. The coefficient of normal restitution for particle-movable piston collisions is in all the
cases $\alpha_{P} =0.99$. It is observed that, for given $\alpha$, there is a region in which $\sigma_{Z} / L$ decreases
as the mass ratio increases. This effect is less pronounced the smaller the coefficient of restitution $\alpha$, i.e.
the more inelastic the collisions. For large values of $M/m$, the results in the figure indicate that $\sigma_{Z}/L$ tends to a plateau with
a constant value. The values of the mass ratio needed to reach the plateau monotonically decrease as $\alpha$
decreases. The data in Fig. \ref{fig3} also indicate that for constant mass ratio, the relative fluctuations increase as $\alpha$
decreases.

\begin{figure}
\includegraphics[scale=0.7,angle=0]{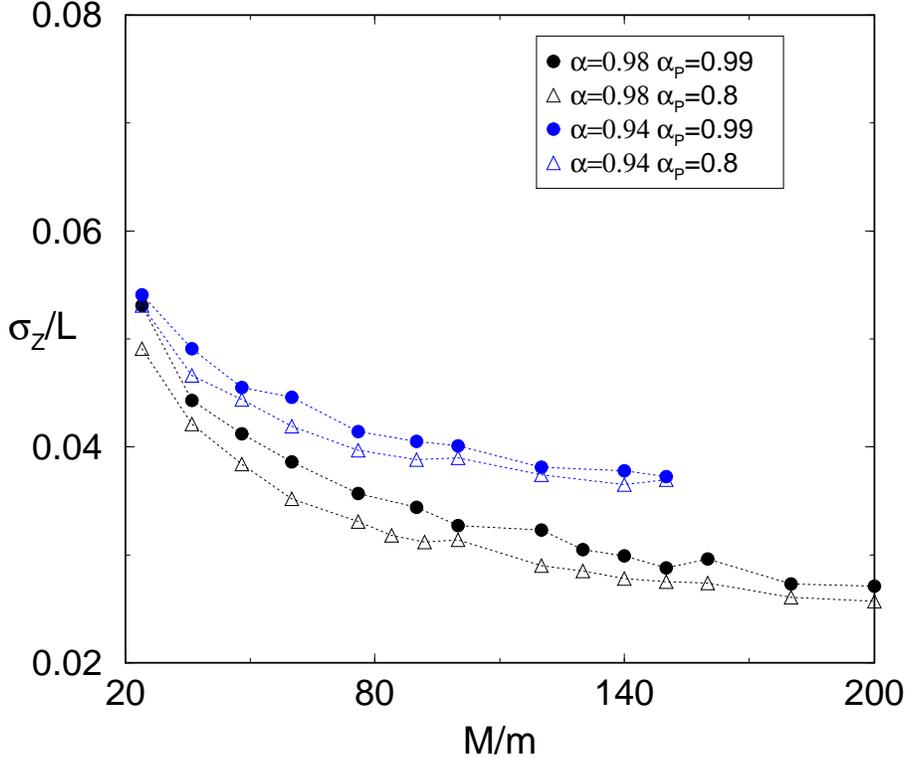}
\caption{(Color online)  Dimensionless relative standard deviation $\sigma_{Z}/L$ of the position of the
movable piston as a function of the mass ratio $M/m$. The symbols are simulation data and the lines guides for the eye.
The two upper curves correspond to $\alpha=0.94$, and the two lower ones to $\alpha =0.98$. In each case,
two values of $\alpha_{P}$ have been employed: $0.99$ (circles) and $0.8$ (triangles).}
\label{fig4}
\end{figure}

To show that the influence of the inelasticity of the particle-movable piston collisions is much weaker than that
of the particle-particle collisions, in Fig. \ref{fig4} the relative standard deviation is plotted as a function
of the mass ratio for two pairs of data. Each pair corresponds to the same value of $\alpha$ (namely, $0.98$ and $0.94$),
but different values of $\alpha_{P}$ (namely, $0.8$ and $0.99$). Although the variation of $\alpha_{P}$ is almost five times
the variation of $\alpha$, it is seen that the data corresponding to the same $\alpha$ are much closer than those
corresponding to the same $\alpha_{P}$. On the other hand, there is a relevant qualitative feature to be stressed.
While decreasing $\alpha$ produces an increase of the relative fluctuations, decreasing $\alpha_{P}$ has the
opposite effect: the relative fluctuations also decrease.

\section{Compressibility and effective temperature}
\label{s3}
\begin{figure}
\includegraphics[scale=0.7,angle=0]{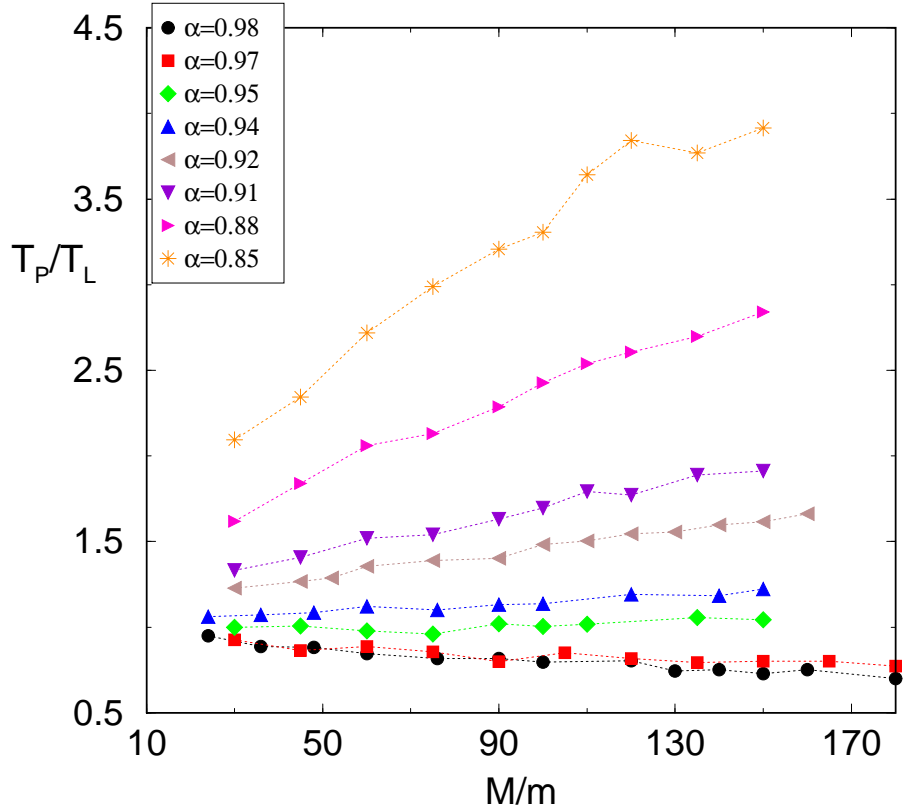}
\caption{(Color online)  Ratio  of the temperature parameter of the piston, $T_{P}$, to the granular temperature
of the gas next to it, $T_{L}$, versus the mass ratio $M/m$, for several values of the coefficient of restitution of the
gas $\alpha$, as indicated in the insert. The curves are guides for the eye. In all the cases, the coefficient of restitution
for the gas-movable piston collisions is $\alpha_{P}=0.99$.}
\label{fig5}
\end{figure}

To measure the facility of the system to be compressed, define a coefficient of compressibility
$k$ by
\begin{equation}
\label{3.1}
k \equiv - \frac{1}{<V>} \left( \frac{\partial <V>}{\partial p_{L}} \right)_{v_{W}},
\end{equation}
where $p_{L}$ is the pressure of the granular gas in the vicinity of the movable piston and $V$ is
the volume (area) of the system. The derivative in the above equation is computed at constant value
of all the parameters defining the system, $\alpha$, $\alpha_{P}$, $N_{z}$ and $v_{W}$, except
$p_{L}=Mg_{0}/W$, as it follows from the definition of the pressure. In the following, $p_{L}$ will be modified by changing the mass of the piston $M$,
keeping  $g_{0}$ and $W$ unchanged. The reason for this choice is twofold.  First, changing $g_{0}$ is equivalent
to modifying $v_{W}$ and, second, increasing $W$ can lead to the set up of transversal instabilities, as already
mentioned. The notation stresses the constancy of the velocity of the vibrating wall. Although other
compressibility coefficients could be defined, the one in (\ref{3.1}) has the
advantage of being easy to implement in experiments, at least at a conceptual level. On the other hand, it is worth
to remark that by keeping $v_{W}$ constant and changing $M$, both the hydrodynamic profiles inside the
fluid and the power injected into it through the vibrating wall are modified. In particular, the latter is given
by \cite{ByR09a}
\begin{equation}
\label{3.2}
Q_{0}= W \left( N_{z}+ \frac{M}{mW} \right) m g_{0}v_{W}.
\end{equation}
Therefore, increasing $M$ while keeping $v_{W}$ constant produces an increase of the power $Q_{0}$.  For the two
dimensional systems being
considered here, the definition (\ref{3.1}) is equivalent to
\begin{equation}
\label{3.3}
k= -\frac{W}{Lg_{0} } \left( \frac{\partial L}{\partial M} \right)_{v_{W}}.
\end{equation}
The idea of introducing a compressibility for vibrated granular fluid was already used in ref. \cite{ACyS02} in the context
of hydrodynamical stability analysis.
In a molecular system at equilibrium, the isothermal compressibility
\begin{equation}
\label{3.4}
k_{T} = -\frac{1}{<V>} \left( \frac{\partial <V>}{\partial p} \right)_{T}
\end{equation}
is related with the volume fluctuations by
\begin{equation}
\label{3.5}
\sigma_{V}^{2} \equiv <(V-<V>)^2> = k_{B}T <V> k_{T},
\end{equation}
where $k_{B}$ is the Boltzmann constant. In non-equilibrium states, there is
no reason to expect the above relationship to hold, but it is tempting to
employ it to define an effective temperature parameter of the system, $T_{eff}$,
expecting it to have some intrinsic physical meaning. Therefore, taking into account
that the volume fluctuations in the case being considered are associated to fluctuations of the
position $Z$ of the movable piston, we define $T_{eff}$ through
\begin{equation}
\label{3.6}
\sigma_{Z}^{2}= - \frac{T_{eff}}{g_{0}} \left( \frac{\partial L}{\partial M} \right)_{{v}_{W}}.
\end{equation}
The Boltzmann constant has been set equal to unity as it is usually done when defining
the granular temperature from the average kinetic energy of the grains. From Eq.\ (\ref{3.6}) it follows that
$T_{eff}$ relates the volume response to a pressure perturbations with the steady volume fluctuations of the
system. Of course, Eq. (\ref{3.6}) by itself is just a mathematical definition and does not add anything to
the physical understanding of the system. On the other hand, the definition would become relevant if
this effective temperature were related to other temperature-like parameters of the system. Two main candidates
clearly stand out: the granular temperature of the gas in the vicinity of the piston, $T_{L}$, and
the temperature parameter of the piston, $T_{P}$, defined through $T_{P} = M <V_{z}^{2}>$ \cite{ByR09a}.
Both parameters, $T_{L}$ and $T_{P}$, are not at all the same, as it should be the case if the system
under consideration were at equilibrium and energy equipartition would apply. Violation of equipartition is
a general feature of granular systems known since long ago \cite{JyM87} and that has attracted a lot of attention
in the last years. A more detailed discussion of this issue for the set up being considered here is given in
\cite{ByR08a}. In Fig. \ref{fig5}, event driven simulation results for the ratio $T_{P}/T_{L}$ are plotted
as a function of $M/m$. Several values of the coefficient of normal restitution of the gas in the interval
$0.85 \leq \alpha \leq 0.98$ have been considered, while again $\alpha_{P}=0.99$ for all the data shown. It is observed that
the behavior of this temperature ratio is quite intricate. For instance, $T_{P}/T_{L}$ decreases as $M/m$ increases
for $\alpha>0.95$, but it happens the other way round for $\alpha < 0.95$. Actually, as $\alpha$ decreases below
this value, the increase of the temperature ratio is rather fast, and $T_{P}$ reaches to be up to four times larger
than $T_{L}$. If the parameters $T_{P}$ and $T_{L}$ were interpreted as real temperature parameters, this latter
behavior would be fully counterintuitive.  The temperature of the heated body (the piston) is larger than the
temperature of the heating one (the gas next to the piston).

\begin{figure}
\includegraphics[scale=0.7,angle=0]{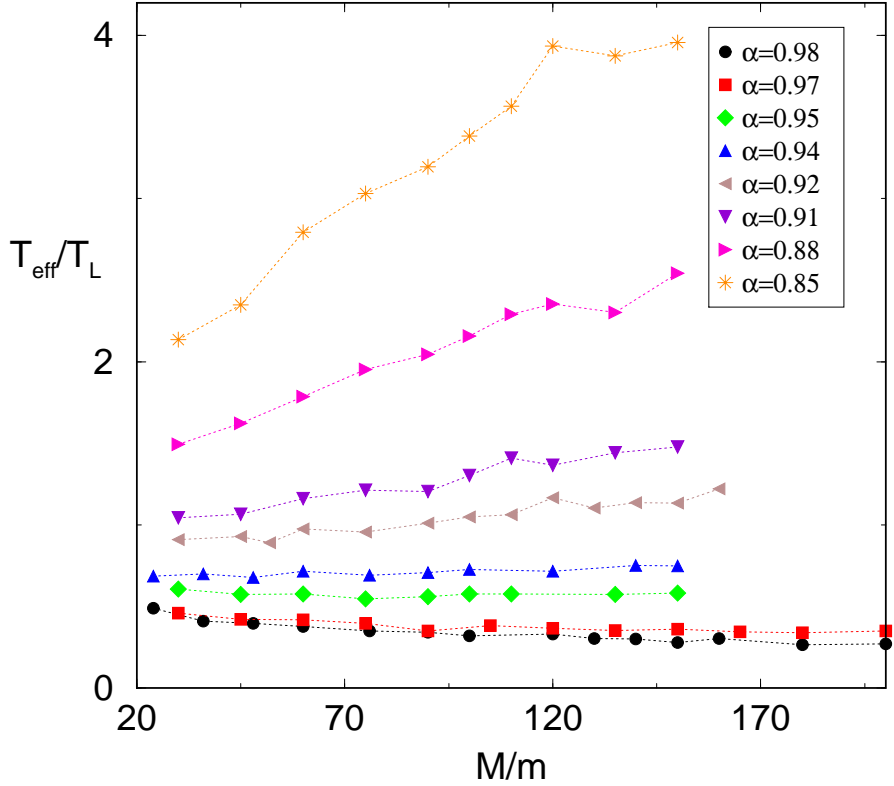}
\caption{(Color online)  Ratio  of the effective temperature, $T_{eff}$, defined in Eq. (\protect{\ref{3.6}}) to the
temperature of the gas in the vicinity of the piston, $T_{L}$, as a function of the mass ratio  for the same systems as in
Fig. \protect{\ref{fig5}}.}
\label{fig6}
\end{figure}

\begin{figure}
\includegraphics[scale=0.7,angle=0]{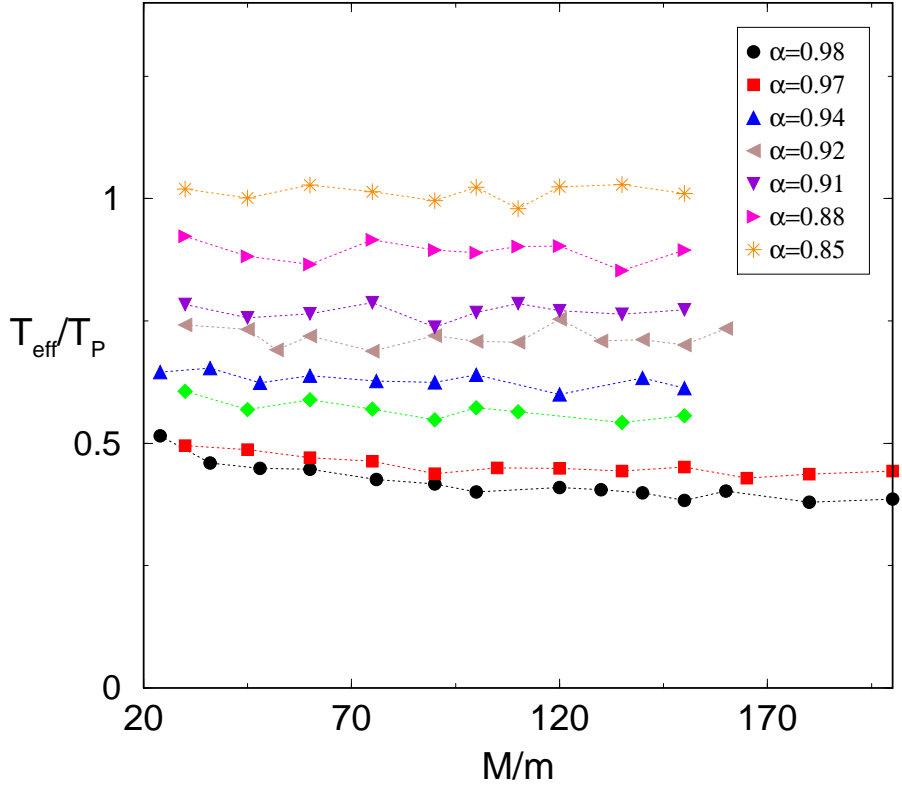}
\caption{(Color online)  Ratio  of the effective temperature, $T_{eff}$, defined in Eq. (\protect{\ref{3.6}}) to the
temperature parameter of the piston, $T_{P}$, as a function of the mass ratio  for the same systems as in
Fig. \protect{\ref{fig5}}. }
\label{fig7}
\end{figure}

From the values of $\sigma_{Z}$ and $(\partial L / \partial M)_{v_{W}}$ obtained from the event driven simulation data, the effective
temperature $T_{eff}$ has been computed by means of its definition, Eq.\ (\ref{3.6}). In the Appendix some details are
given of the way in which the above derivative was actually evaluated. Then, in Figs. \ref{fig6} and \ref{fig7} the
temperature ratios  $T_{eff}/T_{L}$  and $T_{eff}/T_{P}$ are plotted, respectively,  as a function of the mass ratio for the same
systems as in Fig. \ref{fig5}. A clear difference is observed in the behavior of the two temperature ratios for a given
constant coefficient of restitution $\alpha$. While $T_{eff}/T_{L}$ exhibits a strong dependence on $M/m$ and does not
seem to tend to a well defined limit as it increases, the dependence of $T_{eff}/T_{P}$ on $M/m$ is very weak, being only
appreciable for the least inelastic cases and when the mass ratio is small. Thus it is concluded that for
large enough mass ratio $M/m$, the effective temperature is proportional to the temperature
parameter of the piston with a coefficient of proportionality that is independent of the mass of the piston, i.e.
\begin{equation}
\label{3.7}
T_{eff} = b(\alpha,\alpha_{P}) T_{P},
\end{equation}
where a possible dependence of the coefficient on $\alpha_{P}$ has been included.

\begin{figure}
\includegraphics[scale=0.7,angle=0]{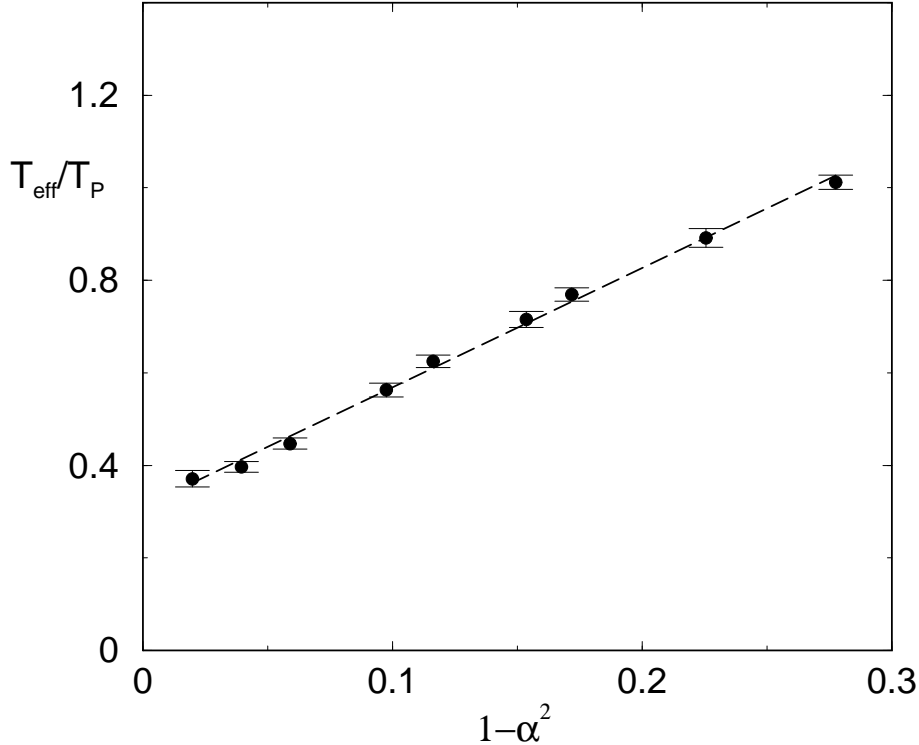}
\caption{Ratio $T_{eff}/T_{P}$ for large values of $M/m$ as a function of $1-\alpha^{2}$ for the same
systems as in Fig. \protect{\ref{fig5}}. The symbols are simulation results and the dashed line a linear fit of them.}
\label{fig8}
\end{figure}

To identify the dependence of the coefficient $b$ on $\alpha$, in Fig. \ref{fig8} the plateau values
of $T_{eff}/T_{P}$, reached upon increasing the value of the mass ratio $M/m$, are plotted versus $1- \alpha^2$.
The points are very well fitted by a straight line, indicating that $T_{eff}/T_{P}$ grows linearly with
$1-\alpha^{2}$, at least in the considered interval, $0.85 \leq \alpha \leq 0.98$. It is worth to note that
some care is needed when extrapolating to the elastic limit $\alpha \rightarrow 1$ these results. In this limit,
a stationary state is only possible if, in addition, the vibrating wall is arrested, i.e. also the limit
$v_{W} \rightarrow 0$ is taken.
But all the previous discussion has been carried out at constant velocity of the vibrating wall.  This explains
why extrapolation of the linear fitting in Fig. \ref{fig8} does not lead to $b=1$, and this does not mean any
kind of contradiction.

\begin{figure}
\includegraphics[scale=0.7,angle=0]{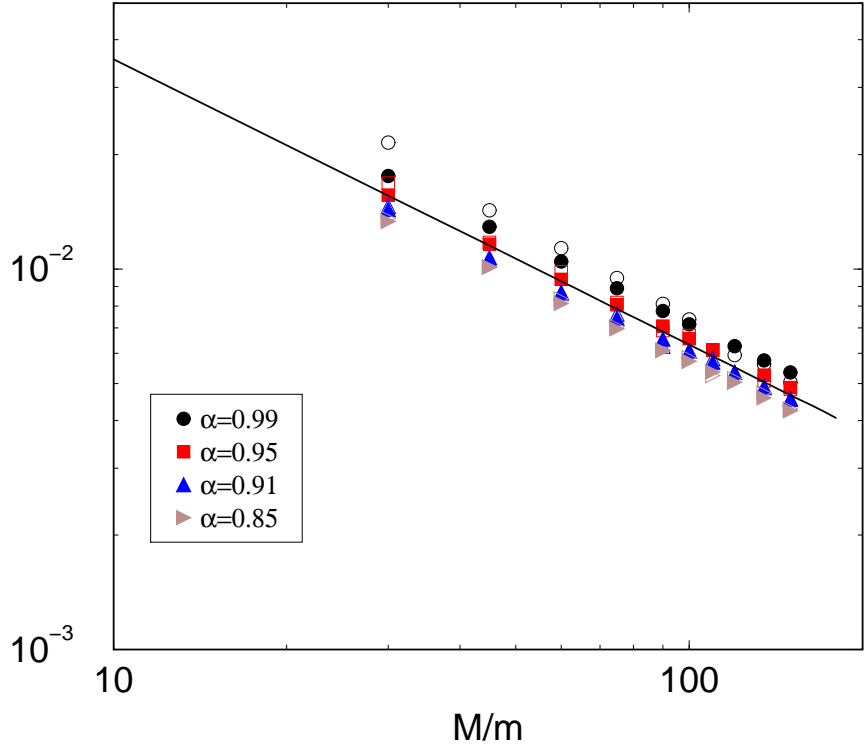}
\caption{(Color online) Dimensionless compressibility $-m L^{-1} \left( \partial L / \partial M \right)_{v_{W}} $ (filled
symbols) and scaled position fluctuations of the piston $\sigma_{Z}^{2} m g_{0} / L b(\alpha,\alpha_{P}) T_{P}$ (
empty symbols) as a function of the mass ratio, $M/m$. The different symbols correspond to different values of the
coefficient of restitution for the collision between particles, $\alpha$, as indicated in the insert. The straight
line has a slope $-3/4$,  and it is a guide for the eye.}
\label{fig9}
\end{figure}

In Fig. \ref{fig9}, the dimensionless compressibility,
$-m L^{-1} \left( \partial L / \partial M \right)_{v_{W}} $ and the scaled second moment of the position fluctuations
$\sigma_{Z}^{2} m g_{0} / L b(\alpha,\alpha_{P}) T_{P}$ have been plotted as functions of $M/m$. A logarithmic
representation is employed. Data for several values of the coefficient $\alpha$ have been included.  For all the
data $\alpha_{P} = 0.99$. If
$T_{eff}$ had been used instead of $ b(\alpha,\alpha_{P}) T_{P}$, the two plotted quantities would be the same
by definition, i.e. the filled and empty symbols would agree in all the cases. It is seen that the dependence on $M/m$
of the dimensionless compressibility is accurately described by a power law of the form $(M/m)^{-3/4}$, indicated in the
figure by the solid straight line.  Nevertheless, it is important to realize that the interval of values of $M/m$
for which the above dependence is identified is rather narrow, just one order of magnitude, so that its range of
validity can be  limited.

\section{Discussion and summary}
\label{s4}
The aim here has been to investigate the volume fluctuations of a vibrated low density
gas of inelastic hard disks in presence of gravity, and confined by a movable piston on the top. The study has
been restricted to the parameter region in which the system reaches a steady state with gradients only in
the direction of the external gravitational field, i.e. perpendicular to the movable piston.
In practice, this has limited the values of the coefficient of normal restitution of the gas particles to
the interval $ 0.85 \leq \alpha < 1$. Due to the coupling between inelasticity and hydrodynamic gradients,
which is peculiar of steady states of granular systems, the above limitation also implies restriction to small gradients.
Nevertheless,  the analysis carried out in ref.\ \cite{ByR09a} indicates that the range of hydrodynamic gradients
considered here exceeds the limit of validity of the Navier-Stokes approximation.

Some of the main results can be summarized as follows: i) for large mass of the movable piston compared with the mass of the gas
particles, the volume fluctuations are Gaussian with very good accuracy, ii) the square
root of the second moment of their distribution scales with the square of the velocity of the vibrating wall at the
bottom, i.e. in the same way as the amplitudes of the hydrodynamic fields in the gas, iii)  by requiring the same relation between volume
fluctuations and compressibility as in equilibrium systems to be verified, an ``effective temperature'' can be defined,
iv) the effective temperature turns out to be proportional to the second moment of the velocity fluctuations of the
piston, with a proportionality parameter that depends on the inelasticity of both the particle-particle and particle-piston
collisions, but it seems to be independent of  the mass of the piston, and v) the effective temperature can not be
related in  a simple way to the temperature of the granular gas; even more, the relationship between both parameters is
not monotonic.

A relevant open question is the relationship between the granular temperature of the gas in the vicinity
of the piston and the temperature parameter of the piston, the latter defined from the second moment of its velocity
distribution. An explanation of the simulation results seems to require a detailed knowledge of the velocity
distribution function of the gas next to the piston \cite{ByR08a}. If this is the case, approximated solutions of the
Boltzmann equation,
as provided by instance by the Chapman-Enskog procedure in the first Sonine approximation, would not
be  of enough accuracy as to describe the deviation from equipartition between the gas and the movable piston.

The present study complements the one in ref. \cite{ByR08a}, in which the velocity fluctuations of the piston were
investigated in detail. A natural issue now is whether the velocity fluctuations and the position
fluctuations of the piston are correlated. We have computed from the simulation data the joint probability distribution
for the position and velocity of the piston and compared it with the product of the marginal
distributions for the position and the velocity. Both results agree within the statistical uncertainties, indicating
the absence of correlations.

\section{Acknowledgements}

This research was supported by the Ministerio de Educaci\'{o}n y
Ciencia (Spain) through Grant No. FIS2008-01339 (partially financed
by FEDER funds).

\appendix

\section{}
\label{ap1}  In ref.  \cite{ByR09a}, an expression for the average position of the piston,
$L$, was derived by using a hydrodynamic description of the granular with the appropriate boundary
conditions. The theoretical prediction was showed to be in reasonable agreement with the simulation
results. Although this expression could have been used to compute  $(\partial L /\partial M)_{v_{W}}$, here this quantity
has been obtained from the simulation data for the sake of consistency.

When trying to compute from the values of $L$ as  a function of $M/m$ the
derivative of the former with respect the latter, the technical problem arises that the considered values of
$M/m$ are separated
by intervals of the order of $10$ or $15$ units. Decreasing this interval would require to strongly increase
the simulation time. In addition, some smoothing process should be used. Here, a different approach
has been followed. The simulation data for $L$ have been fitted by an analytical function of the mass
ratio whose derivative has been afterwards computed. The trivial choice of a polynomial in $m/M$ does not
work so, motivated by the equilibrium elastic result, an expression of the form
\begin{equation}
\label{ap.1}
L = A \frac{g_{0}}{m v_{W}^{2}} \ln \left( 1+ \frac{Bm}{M} \right),
\end{equation}
where $A$ and $B$ are two adjustable dimensionless parameters, was used. It turned out that it fits very well the
results for all the values of $\alpha$ and $\alpha_{P}$ considered here. Moreover, the values of the two fitting
parameters are quite stable, in the sense that their value seem to converge very fast when the number of fitted data
is increased.

\end{document}